\documentclass[conference,a4paper]{IEEEtran}

\usepackage[nolist,nohyperlinks]{acronym} 

\usepackage{xcolor} 
\usepackage{graphicx}
\usepackage{amsmath}
\usepackage{amssymb}

\usepackage{changes}
\usepackage{multicol}

\renewcommand{\baselinestretch}{.952}
\newcommand{\reviewColorText}{black}
\begin{document}
\begin{acronym} 
\acro{AP}{Access Point}
\acro{UE}{User Equipment}
\acro{DL}{downlink}
\acro{UL}{uplink}
\acro{RU}{Radio Unit}
\acro{CU}{Central Unit}
\acro{DU}{Distributed Unit}
\acro{ULA}{Uniform Linear Array}
\acro{CCDF}{Complementary Cumulative Distribution Function}
\acro{CDF}{Cumulative Distribution Function}
\acro{SU}{Single-User}
\acro{MU}{Multi-User}
\acro{MIMO}{Multiple Input Multiple Output}
\acro{SISO}{Single Input Single Output}
\acro{D-MIMO}{Distributed Multiple Input Multiple Output}
\acro{TDD}{Time Division Duplexing}
\acro{OFDM}{Orthogonal Frequency Division Multiplexing}
\acro{CSI}{Channel State Information}
\acro{RSRP}{Reference Received Signal Power}
\acro{ZF}{Zero forcing}
\acro{SVD}{Singular Value decomposition}
\acro{SNR}{Signal-to-Noise Ratio}
\acro{SINR}{Signal-to-Interference-plus-Noise Ratio}
\acro{EE}{Energy Efficiency}
\acro{SE}{Spectral Efficiency}
\acro{LoS}{Line-of-Sight}
\acro{NLoS}{Non Line-of-Sight}
\acro{PE}{Power Efficiency}
\acro{RT}{Ray Tracing}
\end{acronym}
%

\title{Comparative Analysis of Ray Tracing and Rayleigh Fading Models for Distributed MIMO Systems in Industrial Environments}
\author{\IEEEauthorblockN{Aymen~Jaziri\IEEEauthorrefmark{1}, David~Demmer\IEEEauthorrefmark{2}, Yoann~Corre\IEEEauthorrefmark{1},  Jean-Baptiste~Doré\IEEEauthorrefmark{2}, Didier~Le~Ruyet\IEEEauthorrefmark{3},\\ Hmaied~Shaiek\IEEEauthorrefmark{3}, Pascal~Chevalier\IEEEauthorrefmark{3}}
\IEEEauthorblockA{\IEEEauthorrefmark{1}SIRADEL, Saint-Grégoire France, ajaziri@siradel.com}
\IEEEauthorblockA{\IEEEauthorrefmark{2} CEA-Leti, Université Grenoble Alpes, Grenoble, France, david.demmer@cea.fr}
\IEEEauthorblockA{\IEEEauthorrefmark{3} CNAM, CEDRIC Laboratory, Paris, France, didier.le\_ruyet@cnam.fr} 
 \vspace{-0.6cm}}

\maketitle

\begin{abstract}
This paper presents a detailed analysis of coverage in a factory environment using realistic 3D map data to evaluate the benefits of Distributed MIMO (D-MIMO) over colocalized approach. Our study emphasizes the importance of network densification in enhancing D-MIMO performance, ensuring that User Equipment (UE) remains within range of multiple Access Points (APs). To assess MIMO capacity, we compare two propagation channel models: ray tracing and stochastic. While ray tracing provides accurate predictions by considering environmental details and consistent correlations within the MIMO response, stochastic models offer a more generalized and efficient approach. The analysis outlines the strengths and limitations of each model \textcolor{\reviewColorText}{when applied to the simulation of the downlink (DL) and uplink (UL) single-user capacity in various D-MIMO deployment scenarios.}
\end{abstract}

\IEEEpeerreviewmaketitle
\section{Introduction}
\IEEEPARstart{T}{he} introduction of private mobile networks is anticipated to revolutionize various industries, including factory automation, transportation and logistics sectors, significantly influencing their operations. Alongside advancements like Mobile Edge Computing (MEC), Time-Sensitive Networking (TSN), and improved security measures, \ac{D-MIMO} technology is also expected to enhance communication availability and reliability in indoor settings, such as factories. This is achieved by increasing \ac{LoS} coverage and lowering the chances of signal obstruction, as multiple \acp{AP} can serve individual users.


In this paper, we characterize the MIMO propagation channel and present a detailed analysis of the impact of network densification on coverage overlapping and optical visibility.  This is accomplished through the realistic modeling of 3D map data \textcolor{\reviewColorText}{and \ac{RT} simulations} in a factory setting, leveraging the $3.7$~GHz spectrum, which is designed to deliver low-latency, high-capacity private 5G solutions \cite{dore:cea-04213326}. Previous studies have underscored the importance of realistic propagation models in achieving reliable wireless communication in industrial settings \cite{Ozawa18,Greg2024}. In this private network scenario, our objective is to evaluate the potential advantages that \ac{D-MIMO} technology, in conjunction with network densification, can bring to indoor environments. \ac{D-MIMO} systems rely on multiple \acp{AP} to serve \acp{UE}~\cite{DMIMO}. Enhanced coverage overlapping and improved optical visibility ensure that \acp{UE} are within the range of multiple \acp{AP}, facilitating the effective implementation of \ac{D-MIMO}. Additionally, we aim to highlight the disparity between a{ \textcolor{\reviewColorText}{fully \ac{RT}-based} channel model and the Rayleigh-based \textcolor{\reviewColorText}{fading} model that is commonly used in theoretical research. The theoretical benefits of \ac{D-MIMO} and Cell-Free massive MIMO (CF-mMIMO) over conventional co-located MIMO and small non-cooperative base stations have been extensively investigated in the literature. For instance, a comparison between numerous non-cooperative small cells and CF-mMIMO is presented in~\cite{cf}. These studies highlight the potential of \ac{D-MIMO} and CF-mMIMO to provide superior coverage, increased capacity, and enhanced reliability, making them promising solutions for future wireless communication networks. Despite extensive research into \ac{D-MIMO} communications, few studies have evaluated the combined impact of network densification and \ac{D-MIMO} on network performance using realistic \ac{RT} propagation data (e.g. \cite{CFMIMO_raytracing} for mmWave). This work, therefore, contributes to the evaluation of propagation in an industrial environment based on \ac{RT} tools and assesses system performance by incorporating a physical layer model capable of advanced, state-of-the-art processing, such as \ac{UL} cooperative detection and \ac{DL} cooperative beamforming for multi-stream transmissions.

This paper is structured as follows. Section II briefly explains the studied scenario and the considered propagation channel models. Section III details the system model, the power allocation and the spectral efficiency computation. Section IV presents the numerical results and analysis. Finally, Section V concludes the paper and gives future perspective of this work.
\section{System model}

We consider $M \times N$ distributed \ac{MIMO} \ac{OFDM} transmission with $N_c$ sub-carriers.
The \acp{AP} and \acp{UE} operate under a \ac{TDD} protocol, which consists of an \ac{UL} pilot training phase for channel estimation, an \ac{UL} transmission and a \ac{DL} transmission. We assume perfect synchronization among \acp{AP} in terms of time, frequency, and phase. Additionally, channel reciprocity between \ac{UL} and \ac{DL} links is considered, enabling the \ac{MIMO} system to operate in closed-loop mode with \ac{CSI} available at the transmitter. The \ac{UL} channel estimation can thus be used to determine the \ac{DL} precoding coefficients.

The received signal vector $\mathbf{y}_k \in \mathbb{C}^{N\times 1}$ at the $k$-th subcarrier can be written as
\begin{equation}
      \mathbf{y}_{k} = \mathbf{W}_{k}\left ( \mathbf{H}_{k}\sqrt{\mathbf{P}_{k}}\mathbf{F}_{k}\mathbf{Q}_k\mathbf{x}_{k} + \mathbf{n}_{k}\right ) \label{eq:system_model},
\end{equation}
where $\mathbf{x}_{k}\in \mathbb{C}^{N\times 1}$ is the transmitted signal, $\mathbf{n}_{k} \in \mathbb{C}^{N\times 1}$ the noise vector. Each element of the noise vector is uncorrelated and follows a normal distribution, specifically $\mathcal{N}(0,N_0)$. $\mathbf{W}_{k}\in \mathbb{C}^{N \times N}$, $\mathbf{H}_{k}\in \mathbb{C}^{M \times N}$, $\mathbf{P}_{k}\in \mathbb{R}^{M \times M}$, $\mathbf{F}_{k}\in \mathbb{C}^{M \times N}$ and $\mathbf{Q}_{k}\in \mathbb{C}^{N \times N}$ are respectively the linear combining at the receiver side, the propagation channel matrix, the per-antenna amplification gains (diagonal matrix), the linear precoding applied at the transmitter side and the per-layer power allocation (diagonal matrix)\footnote{As channel aging is not considered in the proposed study, we decided to omit the time indices for sake of simplicity.}. In this work, we consider \ac{SU} multi-layer transmissions meaning that independent data streams are transmitted in parallel. Therefore, the precoding/combining techniques aim at isolating the different independent streams and avoid inter-layer interference. Two strategies are considered for \ac{DL}: i) the \ac{ZF} where $\mathbf{F}_k \triangleq \mathbf{H}_k^\dagger$\footnote{where $x^\dagger$ denotes the right Moore-Penrose pseudo inverse of $x$} and $\mathbf{W}_k$ is the identity matrix and ii) the \ac{SVD} where $\mathbf{F}_k \triangleq \mathbf{U}_k^H$ and $\mathbf{W}_k \triangleq \mathbf{V}_k$ for $\mathbf{H_k} = \mathbf{V}^H_k \mathbf{\Sigma}_k\mathbf{U}_k$. The main difference between the two approaches is the presence of a combiner for \ac{SVD}. We assume that combiners $\mathbf{W}_k$ are either supplied to or estimated by the \acp{UE}.
In this scenario, the number of \ac{UE} received antennas matches the number of multiplexed \ac{MIMO} layers. For both techniques, a water-filling algorithm is applied to each carrier. Power allocation is achieved in a way maximizing the achievable capacity \cite{Proakis2007}.

The proposed study aims at comparing the achieved performance of legacy co-localized and distributed architectures.
Indeed, in co-localized scenarios, when the \acp{UE} are served by a unique \ac{AP}, $M$ denotes the number of \ac{AP} antennas. On the other hand, with distributed architectures, $M$ denotes the total numbers of active antennas (\textit{i.e.} number of cooperative \acp{AP} times the number of antenna per \ac{AP}). Indeed, we assume centralized processing, meaning that the \ac{DL} precoding is computed from the \ac{CSI} of all \acp{AP}. This deployment strategy relies on a potential heavy fronthaul use because \ac{DU} and \acp{RU} continuously exchange datas treams and signaling. Local precoding~\cite{Bjornson2020, Interdonato2020} provides better scalability but will not be addressed in the proposed paper and left as a perspective study. In this work, we consider a small-scale deployment (indoor factory) with a limited number of deployed \acp{AP} ($15$ max) which is compatible with centralized processing. Additionally, perfect \ac{CSI} is assumed for theoretical performance evaluation. For the \ac{UL}, we assume no precoding at the UE transmitter side and employ a centralized \ac{ZF} detector for signal detection.
\section{Channel models}
\subsection{Propagation scenario}
For our evaluation and comparison, we use a digital model of a factory area of size $97 \times 36$ $\text{m}^2$. As depicted in Fig.~\ref{fig:FA2}, the area is like a warehouse that is furnished with many metal racks of height $4$ meters. In this scenario, we generate a massive \ac{MIMO} channel samples database. The $15$ considered \acp{AP} are placed at $5$ meters above the ground (just under the ceiling) and $1$ meter away from walls. The $679$ \acp{UE} are uniformly distributed in a grid with a resolution of $2$ meters over the study area, resulting in a density of $0.25$ \ac{UE}/$\text{m}^2$. All users are positioned at height of $1.5$ meters above the ground. Channel predictions are made for \acp{AP} and \acp{UE} equipped with dual-polar (V/H) antenna arrays operating at $3.7$~GHz.

\subsection{Ray tracing model}
The in-factory propagation is first simulated using the Volcano \ac{RT} tool~\cite{Greg2024}, which is capable of handling complex 3D indoor scenarios and efficiently scaling to a large number of UEs. The number of allowed interactions, i.e. two reflections and one diffraction along each ray-path, is a trade-off between computation time and accuracy of the predicted channels that was previoulsy tested in a factory environment~\cite{Bathia2023}. \ac{RT} is used to generate both the average path-loss and the wideband \ac{MIMO} matrix for each \ac{AP}-\ac{UE} link.
As the considered antennas are dual-polarized, a particular attention is given to accurate modelling of the depolarization. A specific calibration work is conducted for adjustment of two correction parameters (a multiplicative factor and an offset) applied to the cross-polarization ratio (XPR) of each interaction. Two reliable sources are considered: the 3GPP InF model \cite{3GPP-InF-model}; and dual-polar measurements recently collected in a machine room \cite{Greg}. In the 3GPP model, a random XPR is applied independently to each ray, with an average of $12$~dB in Line-of-Sight (\ac{LoS}) and $11$~dB in \ac{NLoS}. The standard deviation is $6$~dB for both optical visibilities. As well, the dual-polar measurements indicate an XPR in the range of $10-12$~dB \cite{Greg}. The \ac{RT} calibration is performed such that the corrected XPR's fall into the target range. Note that we also considered a cross-polar discrimination of 20 dB at the transmit and receive antennas, leading to possible additional depolarization. 

\begin{figure}[b]
    \centering     \includegraphics[width=0.42\textwidth]{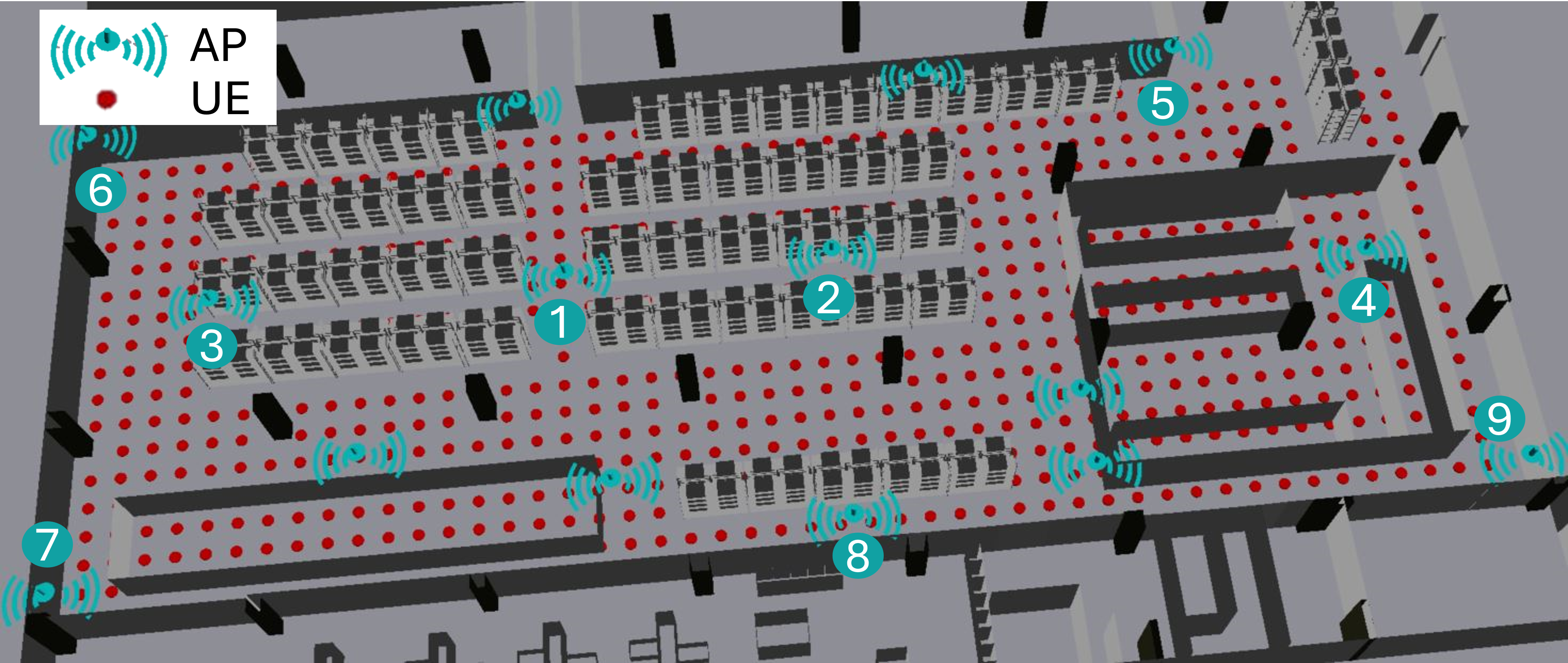}  
    \caption{Study area with the $15$ \acp{AP} and $679$ possible \acp{UE}.}
    \label{fig:FA2}
    \vspace{-0.6cm}
\end{figure}

As the system level simulations are realized for $20$ MHz of bandwidth, we evaluate the channel coherence bandwidth in order to understand the frequency range over which the channel can be considered flat and determine the number of frequencies to be simulated in the synthetic \ac{MIMO} channel matrices. The results from the predictions showed that the coherence bandwidth for 90\% of the considered links with \ac{SISO} path-loss smaller than $170$~dB is higher than $371$~kHz (almost equivalent to 1 resource block (RB) in the 5G NR when the numerology is equal to 1). So, in the channel samples database, a \ac{MIMO} channel matrix is computed for each resource block.

\subsection{Stochastic-based fading model}
To compare with the \ac{RT}-based predictions, we also consider a Rayleigh stochastic model for generation of the \ac{MIMO} channel coefficients. This model is built upon the \ac{RT} path-loss characteristics. Indeed, on average the Rayleigh channels generated for a given link experiences the same channel gains and polarization diversity than in the \ac{RT} prediction. Then, the variation of the channel coefficients along the bandwidth and antenna pairs are randomly selected from a complex normal distribution, scaled by the sub-carrier power model given by \ac{RT}; $\mathbf{H}^{Rayleigh}_k(n,m)=\left|\mathbf{H}^{RT}_k(n,m)\right| \times \mathbf{s}_k(n,m),~\mathbf{s}_k(n,m) \in \mathcal{CN}(0,1)$.
It ensures perfectly independent frequency and spatial channel coefficients; the later characteristics is usually beneficial to the \ac{MIMO} capacity. 
\section{Numerical simulations}
In this section, we use the two channel models to evaluate D-MIMO deployments. First, we present a statistical analysis of the radio propagation conditions met in the considered scenario, based on the  \ac{RT} site-specific prediction for four distinct network designs (indexed in the map Fig. \ref{fig:FA2}):
\begin{multicols}{2}
\begin{itemize}
    \item Single \ac{AP}: Id $1$ 
    \item 3 \acp{AP}: Id $2$,$3$,$4$
    \item $7$ \acp{AP}: Id $1$,$4$,$5$,$6$,$7$,$8$,$9$
    \item $15$ \acp{AP}: All.
\end{itemize}
\end{multicols}

\subsection{Propagation analysis}
We present a variety of propagation coverage performance metrics and analyses, all of which result from the previously mentioned network designs. Our primary objective is to provide a comprehensive overview of the anticipated performance enhancements resulting from network densification, as well as the influence of other network parameters.
\begin{figure}
    \centering
    \includegraphics[width=0.6\columnwidth,trim={0 0cm 0 1.1cm},clip]{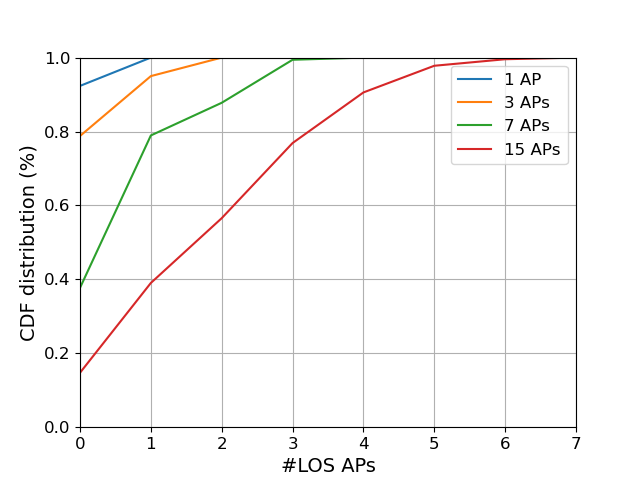}
     \caption{CDF of the number of \ac{LoS} \acp{AP} per \ac{UE}.}
    \label{fig:LOScdf}
    \vspace{-0.5cm}
\end{figure}
In Fig.~\ref{fig:LOScdf}, we plot the \ac{CDF} of the number of \ac{LoS} \acp{AP} to a \acp{UE} of the study area. We observe that with only one \ac{AP} (the blue curve), more than $90\%$ of the \acp{UE} are in \ac{NLoS}. This indicates that a single AP is insufficient to provide reliable LoS coverage to the majority of UEs. However, with the densification of the network, the percentage of \ac{LoS} to one or more \acp{AP} increases significantly. For instance, in the design with 7 \acp{AP}, $40\%$ \acp{UE} are still in \ac{NLoS} from any activated \acp{AP}, $40\%$  \acp{UE} are in \ac{LoS} with one \ac{AP} and $20\%$ \acp{UE} are in \ac{LoS} with at least two \acp{AP}. Furthermore, for the design with $15$ \acp{AP}, more than $40\%$ \acp{UE} are in \ac{LoS} with at least $3$ \acp{AP}. This CDF is a measure of the LoS diversity offered by multi-AP connectivity.

\begin{figure}
    \centering    \includegraphics[width=0.45\columnwidth]{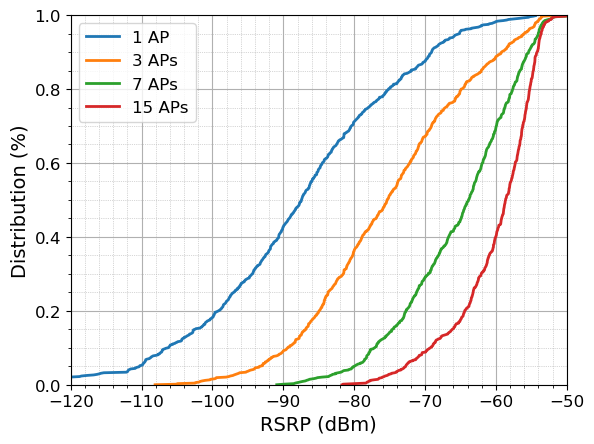}   \includegraphics[width=0.45\columnwidth]{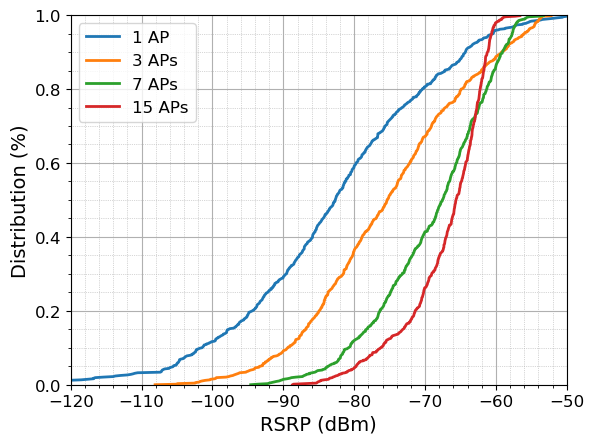}
     \caption{CDF of the best-server \ac{RSRP}, for constant \ac{AP} Tx power (left), or constant network Tx power (right).}
    \label{fig:RSRPcdf}
    \vspace{-0.4cm}
\end{figure}

To further elucidate the benefits derived from network densification, and understand the impact of the transmit (Tx) power distribution among \acp{AP}, we propose an examination of two distinct Tx power models: first, with a constant total Tx power per \ac{AP} \textit{i.e.} $23$~dBm for $20$~MHz bandwidth; second, with a constant Tx power for the entire network \textit{i.e.} $27.8$~dBm (equivalent to the sum of 3 \acp{AP} in the previous approach). In Fig.~\ref{fig:RSRPcdf}, we plot the \ac{CDF} of the Reference Signal Received Power (\ac{RSRP}) from the best serving \ac{AP} to each \ac{UE}. As expected, when \ac{AP} Tx power is constant, the network densification  significantly enhances the best-server \ac{RSRP}. However, the improvement gap narrows when considering more than $7$ \acp{AP}. At constant  network Tx power, the  densification leads to a smaller but still marked \ac{RSRP} improvement, especially at lower levels. The impact on highest \ac{RSRP} levels, typically above $-65$~dBm, may be negligible, even negative. Whatever the Tx power strategy, the densification promotes more uniform coverage performance, as indicated by the narrower \ac{RSRP} distribution.

\begin{figure}
    \centering
    \includegraphics[width=0.47\columnwidth]{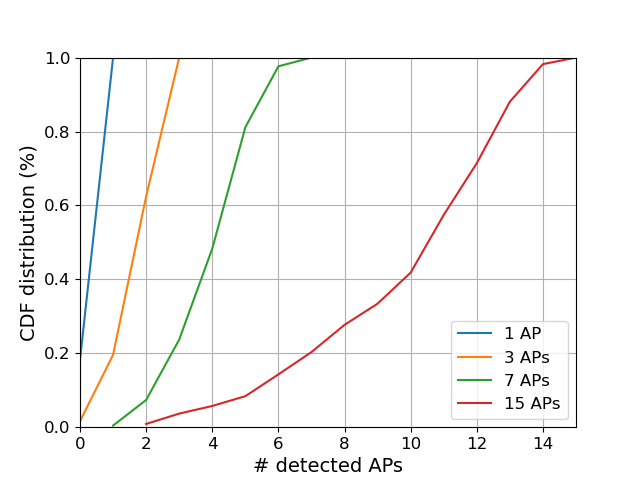}
    \includegraphics[width=0.47\columnwidth]{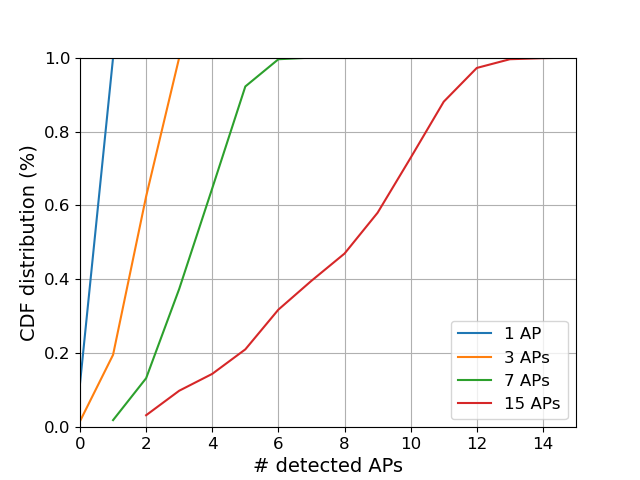}
     \caption{CDF of the number of detected \acp{AP}, for constant \ac{AP} Tx power (left), or constant network Tx power (right).}
    \label{fig:overlap}
    \vspace{-0.5cm}
\end{figure}
Fig.~\ref{fig:overlap} shows the distribution of the number of detected \acp{AP} per \ac{UE} considering an \ac{RSRP} threshold of $-100$~dBm. For both Tx power models, the densification allows to significantly improve the coverage (at least one \ac{AP} detected). More than $80\%$ \acp{UE} detect at least $2$ different \acp{AP} for both Tx power models when the network is composed of at least $7$ \acp{AP}. These results have significant impact on \ac{D-MIMO} systems, which leverage multiple \acp{AP} to serve \acp{UE} simultaneously, exploiting spatial diversity and reducing interference. The performance depends on \acp{AP} overlap, thus on both the network density and power distribution strategy.

\begin{figure}
    \centering
    \includegraphics[width=0.6\columnwidth]{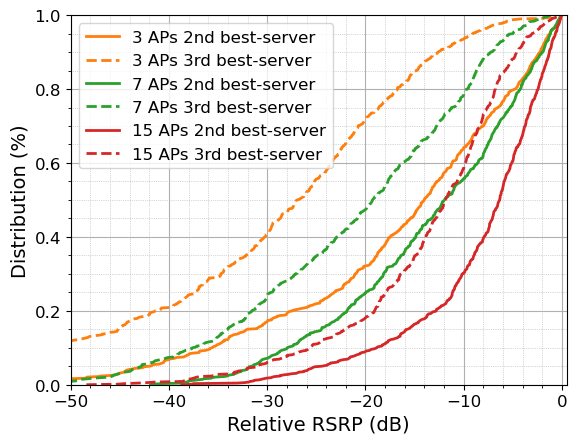}
     \caption{Distribution of the relative \ac{RSRP} for second and third best-serving \acp{AP}. }
    \label{fig:relativeRSRP}
    \vspace{-0.1cm}
\end{figure}

Fig.~\ref{fig:relativeRSRP} illustrates the relative \ac{RSRP} of the 2nd and 3rd best-serving \acp{AP} compared to the highest \ac{RSRP}. Close to 0 dB, the spatial diversity offered by the multi-AP connectivity is more likely able to provide several parallel streams of equivalent capacity. This evaluation helps to determine the suitability of deploying \ac{D-MIMO} in a given scenario. As anticipated, the relative \ac{RSRP} decreases significantly with densification. For instance, the relative \ac{RSRP} of the 2nd best-server is above $-10$~dB for about $70\%$ \acp{UE} when considering maximum densification, but for only $45\%$ with $7$ \acp{AP}. The impact is even more pronounced for the third best \ac{AP}.

\begin{figure}
    \centering
    \includegraphics[width=0.45\columnwidth]{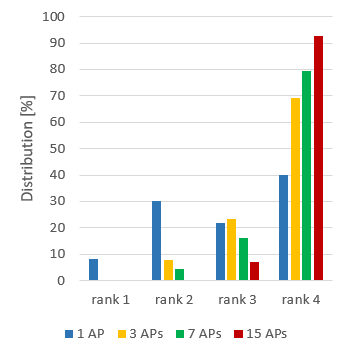}
    \includegraphics[width=0.45\columnwidth]{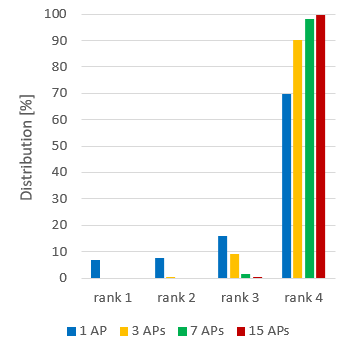}
     \caption{MIMO channel rank distribution for \ac{RT} prediction (left) and Rayleigh distribution (right).}
    \label{fig:rank}
    \vspace{-0.5cm}
\end{figure}

The above results are valid for both the \ac{RT} and stochastic channel models, since only the average power was considered. However, that changes when dealing with the distribution of the \ac{MIMO} channel ranks, as presented by Fig.~\ref{fig:rank}. The channel rank corresponds to the number of significant streams (\textit{i.e.} with a received power greater than $-100$~dBm/RB) assuming a uniform power allocation of $17$~dBm for the $4$ antenna elements at the \ac{AP} side. The channel rank is evaluated for  the best-server only of each \ac{UE} according to the different \ac{AP} groups. First, one can observe that most of the channels admits a rank greater or equal to $2$ thanks to the polarization diversity. Besides, densifying the network improves the channel matrix rank as it increases the stream powers by reducing the \ac{AP}-\ac{UE} distance. Regarding Rayleigh distribution, it provides the best proportion of full-rank matrices thanks to the absence of spatial correlation and therefore enhances link channel diversity (except for single-AP poor-quality links at rank 1). 

Based on this characterized propagation environment, we are now studying the capacity of DL and UL transmissions.

%
%
%

\subsection{UL Cooperative detection}
We evaluate the performance of multi-layer \ac{SU}-\ac{MIMO} transmissions for the \ac{UL} with $52$ RBs. It's worth mentioning that \ac{RT} provides a model at the RB scale. Consequently, simulations are conducted at the RB level, while performance measurements (capacity) are provided at the sub-carrier level (where 12 sub-carriers constitute one RB). The \ac{UE} transmits $4$-layers with a total power of $23$~dBm, \textit{i.e.} $17$~dBm per antenna. Besides, $4$-element \acp{ULA} are considered for each \ac{AP} and for \ac{UE}. Among the $4$ antennas, $2$ are H-polarized, and the other $2$ are V-polarized. The co-polarized antennas are separated by $\lambda/2$ (where $\lambda$ is the wavelength). The cross polarization between antennas is simulated with a cross polar discrimination equal to $20$~dB. For illustration, Fig.\ref{fig:capacity_map_UL} presents the capacity map for quad-layer transmissions, \ac{RT} channel model and four deployments ($1$, $3$, $7$ and $15$ \acp{AP}). Since graphical representations can be challenging to compare across different configurations, a statistical analysis is preferred.

\begin{figure}
    \centering
    \includegraphics[width=1\columnwidth,trim={0 0.2cm 0 0cm},clip]{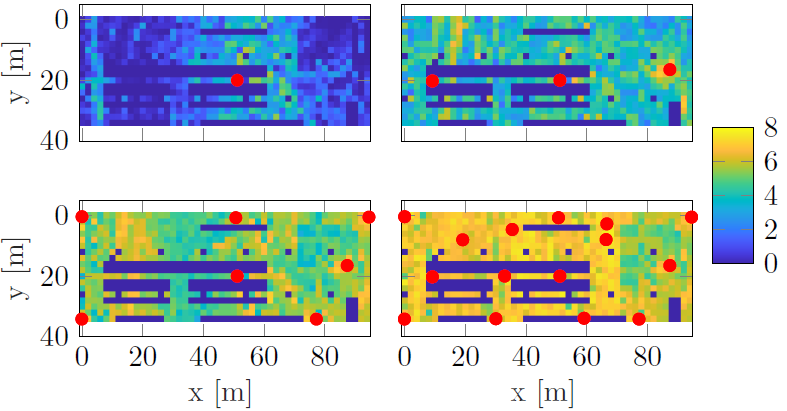}
    \caption{Capacity [bits/s/Hz] map for UL 4-layer transmissions and the $4$ deployment scenarios ($N_0 = -118$~dBm/RB).}
    \label{fig:capacity_map_UL}
    \vspace{-0.5cm}
\end{figure}

Fig.~\ref{fig:capacity_UL} shows the \ac{CCDF} of the capacity per sub-carrier for both the \ac{RT} and Rayleigh modes with different levels of cooperation. The configuration are labeled $(a,b)$: $b$ \acp{AP} are active among $a$ possible. The selection of which \acp{AP} to use is based on the strength of the received signal power. This criterion is advantageous due to its simplicity. However, other criteria, such as those based on the rank of the channel matrix, can be considered but would introduce additional computational complexity.
\begin{figure}
    \centering
    \includegraphics[width=0.49\textwidth]{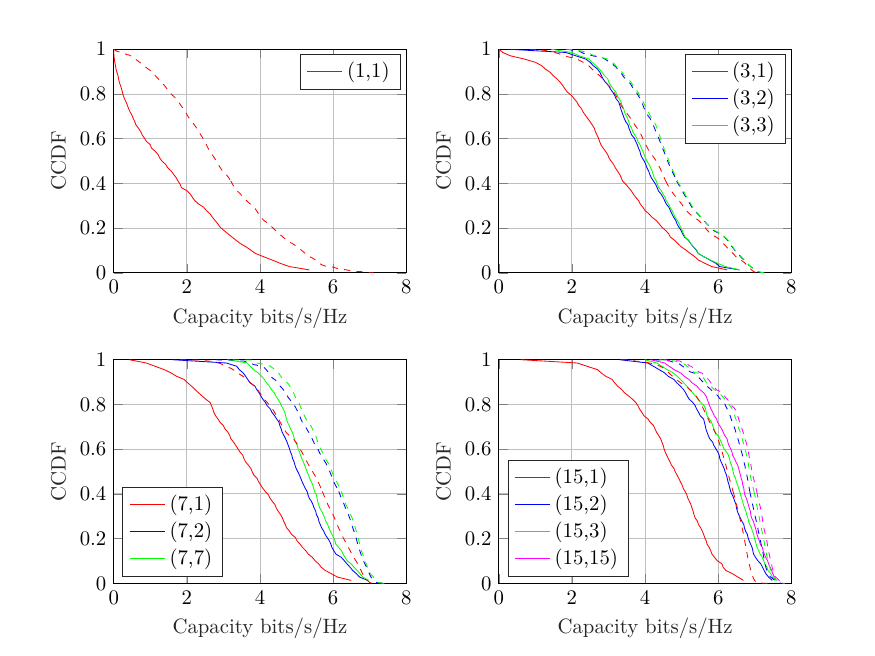}
    \caption{\ac{CCDF} of the capacity for $4$-layers transmission. The dotted lines represent the Rayleigh channel model, while the solid lines correspond to the \ac{RT} model ($N_0 = -118$~dBm/RB).}
    \label{fig:capacity_UL}
    \vspace{-0.5cm}
\end{figure}
In the single \ac{AP} scenario, the Rayleigh model, being richer in diversity, demonstrates better performance. In the case of cooperation, we clearly observe the benefit of distributing antennas to exploit diversity. There is a threshold effect: beyond a certain number of cooperating \acp{AP}, no further diversity gain is achieved. Since the selection of the number of cooperating \acp{AP} is based on the received power metric, increasing cooperation beyond $3$ \acp{AP} brings limited benefit. This is because we reach a certain level of diversity, and the focus is primarily on users close to the \acp{AP}, resulting in channels that are predominantly \ac{LoS} and have a good \ac{SNR}. The Rayleigh model shows the same trends for this scenario and remains optimistic compared to \ac{RT}.

%
%
%
\subsection{DL Cooperative single user beamforming}


We evaluate the performance of multi-layer \ac{DL} \ac{SU}-\ac{MIMO} transmissions. It means that for each time/frequency resource only one \ac{UE} is served by several independent datastreams ($2$ or $4$ in this evaluation). 

For the following numerical simulations, we consider a total transmit power fixed to $23$~dBm to be spread over all available \acp{AP} and antennas. As a consequence, varying the number of active \acp{AP} changes the number of active antennas but keeps the same transmit power at the network level equal to $23$~dBm. The \ac{AP} antenna array is the same one considered for the \ac{UL} study. The \acp{UE} are equipped with $2$ and $4$ antennas, thus are able to receive $2$ (resp. $4$) symbols streams (layers).

The evaluation of the median capacities is given in Fig.\ref{fig:median_capacity_raytracing}. In two-layer transmissions, densifying the network increases the achievable capacity. Besides, \ac{AP} cooperation is beneficial when a low number of serving \acp{AP} is considered. Indeed, we observed with Fig.\ref{fig:rank} that most of the predicted channel matrix admit a rank of at least $2$ which corresponds to the two polarizations. There is thus no need of spatial diversity which explains why joint transmission does not provide better results. This analysis is valid for both the ZF and SVD precoders and both the \ac{RT} and Rayleigh channels. It seems worth noticing that the capacity does not decrease with the number of serving \acp{AP}, and thus is not degraded due to the reduced power per antenna, thanks to the sum-power constraint considered at the network level.

When it comes to the quad-layer transmissions, one can also observe that densifying the network improves the network capacity. \ac{AP} cooperation can also enhance the capacity especially for \ac{RT} because there are less full-rank channel matrices with one serving \ac{AP} as observed with Fig.~\ref{fig:rank}. There is then a need of extra spatial diversity to properly multiplex the $4$ datastreams. It is achieved with a limited number of serving \acp{AP}. In terms of precoding techniques, \ac{SVD} precoding offers better performance compared to \ac{ZF}, mainly due to the effectiveness of the applied postcoder/combiner.


\begin{figure}
    \centering
    \includegraphics[width=0.9\columnwidth]{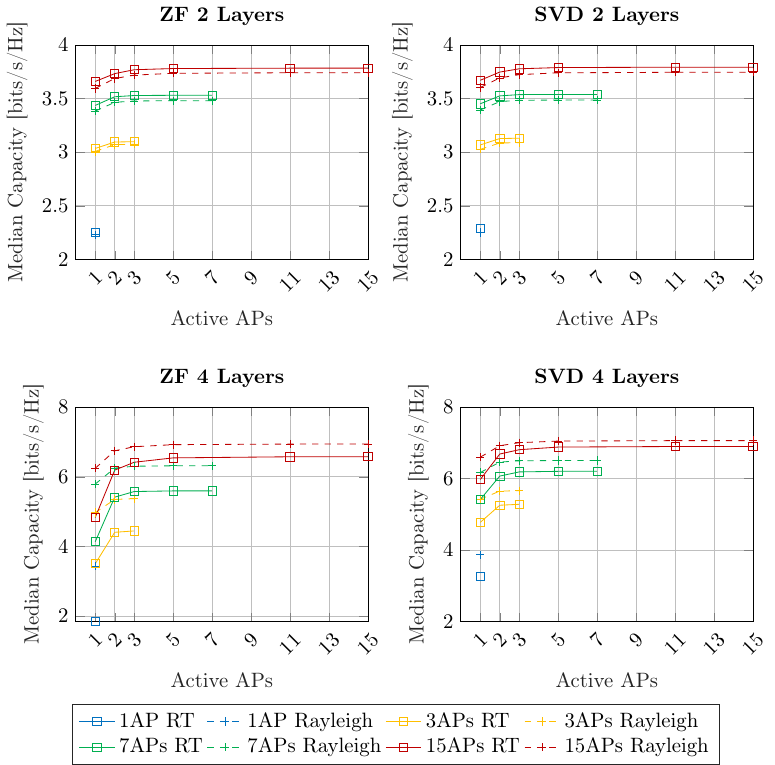}
     \caption{Median capacity for different configuration and \ac{RT} channel model with $N_0 = -118$~dBm/RB. }
    \label{fig:median_capacity_raytracing}
    \vspace{-0.5cm}
\end{figure}

\section{Conclusion}

In this paper, we have presented a comprehensive analysis of coverage and capacity results in a single-user factory scenario using \ac{D-MIMO} concept and realistic RT-based path-loss (including polarization diversity). The spatial fading component is derived either from RT or a stochastic Rayleigh model. The overall performance and behavior trends are quite similar across both channel models. However, the stochastic approach is found to be optimistic compared to RT, especially when no spatial correlation is considered which is frequently assumed in most D-MIMO numerical studies. Besides, network densification significantly enhances performance in both \ac{UL} and \ac{DL}. Although densification and cooperation are distinct concepts, effective densification combined with limited cooperation, involving a small number of \acp{AP}, is often sufficient to substantially optimize performance.  It also emphasizes the crucial role of clustering strategies in improving overall network performance.
\section*{Acknowledgment}
 Part of this work received funding from the French National Research Agency within the frame of the project POSEIDON (ANR-22-CE25-0015).
\bibliographystyle{IEEEtran}

\bibliography{IEEEabrv,bibliography}
\end{document}